\documentclass[aps,prd,twocolumn,preprintnumbers,superscriptaddress,nofootinbib]{revtex4-1}
\usepackage{graphicx}
\usepackage{epstopdf}
\usepackage{amsmath}
\usepackage{amsfonts}
\usepackage{amssymb}
\usepackage{appendix}
\usepackage{enumerate}
\usepackage{natbib}
\usepackage{comment}
\usepackage{bbold}
\usepackage[shortlabels]{enumitem}
\usepackage{color}
\usepackage{slashed}
\usepackage{subfigure}
\usepackage{setspace}
\usepackage{footnote}
\usepackage{lipsum}
\usepackage{multirow}
\usepackage{float}
\usepackage[colorlinks = true,
            linkcolor = blue,
            urlcolor  = blue,
            citecolor = blue,
            anchorcolor = blue]{hyperref}
\usepackage[capitalize]{cleveref}
\usepackage{braket}
\usepackage{multirow}
\usepackage{physics}
\usepackage{feynmp-auto}
\usepackage[normalem]{ulem}
\usepackage{url}
\usepackage{units}

\newcommand{\be}{\begin{eqnarray}}
\newcommand{\ee}{\end{eqnarray}}
\newcommand{\ba} {\begin{equation}\begin{aligned}}
\newcommand{\ea} {\end{aligned}\end{equation}}
\newcommand{\bg} {\begin{equation}\begin{gathered}}
\newcommand{\eg} {\end{gathered}\end{equation}}

\newcommand{\beq}{\begin{equation}}
\newcommand{\eeq}{\end{equation}}

\usepackage{tikz,xcolor,hyperref}

\definecolor{lime}{HTML}{A6CE39}
\DeclareRobustCommand{\orcidicon}{\hspace{-1mm}
	\begin{tikzpicture}
		\draw[lime, fill=lime] (0,0) 
		circle [radius=0.12] 
		node[white] {{\fontfamily{qag}\selectfont \tiny \,ID}};
		\draw[white, fill=white] (-0.0525,0.095) 
		circle [radius=0.007];
	\end{tikzpicture}
	\hspace{-3mm}
}

\foreach \x in {A, ..., Z}{\expandafter\xdef\csname orcid\x\endcsname{\noexpand\href{https://orcid.org/\csname orcidauthor\x\endcsname}
		{\noexpand\orcidicon}}
}

\begin{document}

\title{ Stability and Structural Properties of Hot Quark Stars within Perturbative QCD}% Force line breaks with \\

\author{Tousif Raza\orcidA}
\email{tousif.m.raza@gmail.com}

\affiliation{Department of Physics, Oklahoma State University, Stillwater, OK, 74078, USA}

\author{Tyler Gorda\orcidB}

\email{gorda.1@osu.edu}

\affiliation{ Center for Cosmology and AstroParticle Physics (CCAPP), Ohio State University, Columbus, OH 43210}

\affiliation{ Department of Physics, The Ohio State University, Columbus, OH 43210, USA}

\date{\today}% It is always \today, today,

\begin{abstract}
The hypothesis of strange quark matter (SQM) and the possible existence of strange stars have been extensively investigated within the thermodynamic bag model, typically employing the free Fermi gas approximation with or without perturbative QCD corrections at zero temperature, where quark confinement is modeled by the bag pressure. In this work, based on the perturbative inclusion of quark mass effects, we develop a thermodynamically consistent equation of state (EOS) for SQM at finite temperature, incorporating perturbative corrections up to $\mathcal{O}(\alpha_s)$ in the strong coupling constant $\alpha_s$ while ensuring full adherence to the Maxwell relations. These corrections are essential for accurately describing hot quark stars potentially formed during core-collapse events. We present results for both fixed and running couplings, using a phenomenological model to include the breakdown of the perturbative running of $\alpha_s$ at low momenta. Our results demonstrate that incorporating finite-temperature perturbative QCD corrections leads to SQM configurations that fall within the absolute stability window, with equilibrium energies per baryon lying below that of iron, the most tightly bound nucleus. Our EOS supports compact stars with masses exceeding $1.4\,M_\odot$ (and above $2\,M_\odot$ for some values of the bag constant), in agreement with current astrophysical constraints from pulsar and gravitational-wave observations.

\end{abstract}

\maketitle

\section{Introduction}

The properties of strange quark matter (SQM) have attracted considerable attention since  Witten's seminal proposal that it might constitute the absolute ground state of hadronic matter~\cite{Witten:1984rs}. At zero temperature and pressure, the ground state of ordinary hadronic matter is characterized by~$^{56}\mathrm{Fe}$ the most tightly bound atomic nucleus which  possesses an energy per baryon~${E/A \approx 930.4~\mathrm{MeV}}$. Witten's hypothesis that the energy per baryon of SQM could fall below that of iron motivated  Farhi and Jaffe~\cite{Jaffe} to systematically investigate the stability condition,  $(E/A)_{\mathrm{SQM}} < (E/A)_{{}^{56}\mathrm{Fe}}$. 
Within the MIT bag-model framework,  they explored the dependence of this condition on the strange quark mass~$m_s$ and the  bag constant~$B$, identifying a distinct ``stability window'' in the $m_s$--$B$ plane  where SQM is energetically favored. This stability window provides a strong theoretical  motivation for exploring the astrophysical manifestations of stable strange matter in  ultra-dense environments.

Quark matter is generally expected to become thermodynamically favorable only under extreme conditions, such as those realized in heavy-ion collisions or within the ultra-dense cores of neutron stars. 
The astrophysical consequences of stable SQM, particularly the potential existence of strange quark stars, were subsequently explored by Haensel et al.~\cite{Haensel:1986qb} and Alcock et al.~\cite{Alcock}. 
Their investigations demonstrated that such self-bound stars could closely resemble traditional neutron stars in their macroscopic observables, including mass and radius. 
Strange quark stars have been extensively studied at zero temperature~\cite{Madsen:1998uh,1999JPhG...25..195W, Fraga:2001id,Bombaci_2005,Weissenborn:2011qu,Kurkela:2014vha, Yang:2023haz, Chen:2006zc, Fraga:2022yls, Burgio:2001mk,Schertler:1996tq, TOvergard_1991, Weissenborn:2011qu, Alford:2004pf,Chakrabarty:1989bq,Chakrabarty:1991ui, Prasad:2003bw,Annala:2019puf,Zhang:2024xod, PeresMenezes:2005atu}, and descriptions of the finite-temperature equation of state (EOS) for quark matter have also been discussed across various contexts~\cite{MULLER1981111,REINHARDT1988133,Kettner:1994zs,Bordbar:2011qx,Burgio_2016, PhysRevC.105.045806,Chu:2021aaz}. 

Despite these significant developments, the combined effects of finite
temperature and perturbative QCD (pQCD) corrections have received
comparatively limited attention within a unified EOS framework. The
present work addresses this gap by developing a thermodynamically
consistent treatment of SQM at finite temperature that systematically
incorporates perturbative corrections up to
$\mathcal{O}(\alpha_s)$.

As a recent example, Grippa et al.~\cite{Grippa:2024ppo}, in their study
of the general-relativistic signatures of strange stars, employ an EOS
obtained through a minimal finite-temperature extension of the
zero-temperature result. We identify the following limitations of that prescription:
\begin{itemize} 
    \item \textbf{Inconsistent temperature treatment:} 
    The free-quark contribution to the grand potential includes a finite-temperature correction, while the $\mathcal{O}(\alpha_s)$ correction is added as a separate term evaluated exclusively in the zero-temperature limit. 
    \item \textbf{Inconsistent mass treatment:} The grand chemical potential for free quarks is computed assuming a nonzero quark mass, while the $\mathcal{O}(\alpha_s)$ perturbative corrections are evaluated assuming massless quarks. 
\end{itemize} 
The impact of these limitations can be illustrated through a straightforward exercise: evaluating the Fermi--Dirac integrals for the
pressure and energy density of a free quark gas at fixed nonzero temperature, both in the massless limit and for finite quark masses. The comparison shows that finite-mass corrections can appreciably modify the equation of state, particularly when the quark mass is not negligible compared with the relevant temperature or chemical-potential scale. The importance of finite quark-mass effects in dense quark matter has also been emphasized in Ref.~\cite{Fraga:2004gz}.
A reliable description must therefore incorporate finite quark masses and perturbative corrections within a unified, thermodynamically consistent framework. In this work, we develop such a framework and obtain the finite-temperature equation of state of strange quark matter within pQCD.

The remainder of this paper is organized as follows. 
Section~\ref{sec:Strange Quark Matter with Perturbative QCD} presents our finite-temperature thermodynamic treatment of SQM within perturbative QCD, detailing the resulting EOS and its stability conditions. 
Section~\ref{sec:Global Properties of Quark Stars} examines the global properties of the corresponding quark stars, focusing specifically on mass--radius relations, tidal deformability, and compactness. 
Finally, we summarize our primary findings and conclude in Section~\ref{sec:conclusion}.

\section{Strange Quark Matter in Perturbative QCD at Finite Temperature}
\label{sec:Strange Quark Matter with Perturbative QCD}
\label{sec:intro}
\begin{figure*}
    \centering
    \includegraphics[width=0.45\linewidth]{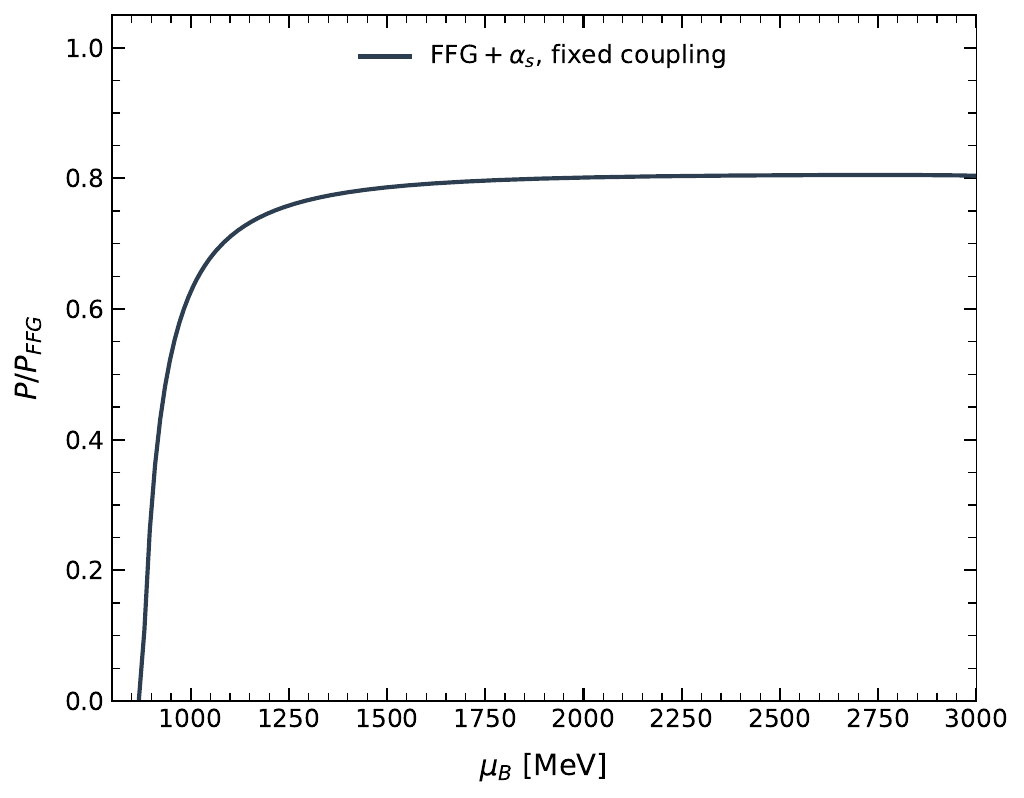}
    \includegraphics[width=0.45\linewidth]{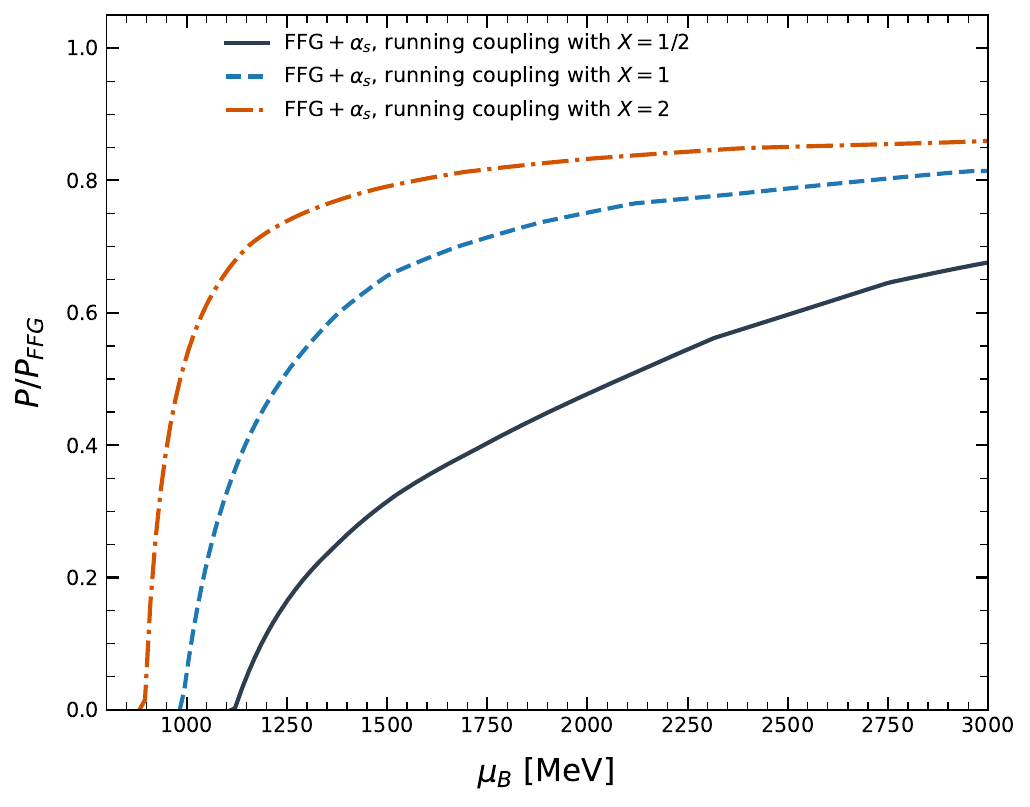}
    \caption{Pressure of strange quark matter as a function of the baryon chemical potential, computed using a fixed QCD coupling, $\alpha_s \approx 0.3$ (left panel), and a running QCD coupling (right panel). In both panels, we set the bag constant to $B^{1/4}=145~\mathrm{MeV}$ and use the quark masses $m_u=m_d=0$ and $m_s=96~\mathrm{MeV}$.}

    \label{fig:pres1}
\end{figure*}

We formulate the equation of state of strange quark matter within
perturbative QCD, considering both massless and finite-mass strange
quarks at nonzero temperatures. The zero-temperature limit can also be
readily recovered within our framework. In this section, we present the
details of the EOS formulation.

\subsection{Massless Strange Quark Matter in Perturbative QCD}

We begin with the simple MIT bag model, in which quarks are treated as free particles confined within a finite region of space and forbidden to escape. 
All long-range, non-perturbative effects of QCD are encoded in a single phenomenological parameter the bag constant $B$, which represents the vacuum pressure, the energy-density difference between the physical vacuum inside the bag and the perturbative vacuum outside. 
In this picture, the vacuum pressure acts as an external confining force that holds quark matter together, providing a crude but effective description of the confinement mechanism inherent in QCD~\cite{Johnson:1975zp, DeTar:1983rw, Chodos:1974je}. 
Despite its simplicity, the bag model has proven remarkably useful in exploring the macroscopic properties of quark matter and remains a valuable tool for generating qualitative and semi-quantitative predictions for strange stars and other exotic compact objects.

In the following, we first discuss the thermodynamics for massless quarks to establish the baseline formalism. 
The massless limit, while idealized, is instructive as it allows for simpler analytic treatment and provides clear insight into the temperature and density dependence of the thermodynamic quantities.\footnote{When a finite quark mass is later introduced, the corresponding charge neutrality and beta equilibrium conditions must be modified accordingly due to the mass-dependent contributions to the number densities and chemical potentials.} 
In the massless approximation, the thermodynamic potential in the grand canonical ensemble $\Omega(m=0,\mu,T)$, including perturbative corrections up to $\mathcal{O}(\alpha_s)$, can be written as follows~\cite{Vuorinen:2003fs,Kurkela:2016was}:
\begin{equation}
    \Omega = \Omega_1(m_s=0,\mu,T) + \alpha_s \Omega_2(m_s=0,\mu,T),
\label{eq:massless potential}
\end{equation}
where the non-interacting contribution is given by
\begin{equation}
    \Omega_1 := -\frac{\pi^2}{45}\frac{T^4}{N_f}\sum_f \left\{ d_A + \left(\frac{7}{4} + 30\bar{\mu}_f^2 + 60\bar{\mu}_f^4\right) d_F \right\},
\end{equation}
and the $\mathcal{O}(\alpha_s)$ correction reads
\begin{equation}
    \Omega_2 := \frac{d_A}{144}\frac{T^4}{N_f}\sum_f \left\{ C_A + \frac{T_F}{2}\left(1 + 12\bar{\mu}_f^2\right)\left(5 + 12\bar{\mu}_f^2\right) \right\}.
\end{equation}

% Converted the g^2 to a 4 Pi alpha_s and put the 4 Pi here.
%
Here, $d_A := N_c^2 - 1$ and $d_F := N_c N_f$ are the dimensions of the adjoint and fundamental representations of the color gauge group, respectively, $T_F := N_f/2$, and
\begin{equation}
    C_F := \frac{N_c^2 - 1}{2N_c}
\end{equation}
is the quadratic Casimir invariant of the fundamental representation. 
The dimensionless chemical potential is defined as
\begin{equation}
    \bar{\mu}_f := \frac{\mu_f}{2\pi T},
\end{equation}
where $\mu_f$ is the chemical potential for quark flavor ${f = u, d, s}$. The summation over flavors accounts for the contributions of each quark species, weighted by their respective degeneracy factors.

Including the bag contribution, the pressure in the grand canonical ensemble is given by
\begin{equation}
    P = -\Omega - B.
\end{equation}
The corresponding energy density $\varepsilon$ is obtained from the thermodynamic relation
\begin{equation}
    \varepsilon = \Omega + \sum_i \mu_i n_i + T s + B,
\end{equation}
where ${s := -\partial \Omega/\partial T}$ is the entropy density,
and ${n_i := -\partial \Omega/\partial \mu_i}$ is the net number density of particle species $i$.

To determine the thermodynamic state of the system, we must specify the chemical potentials for each quark flavor. 
In compact-star matter, the system is expected to be in chemical equilibrium under weak interactions and must satisfy local charge neutrality. 
Imposing these conditions allows us to solve for the quark chemical potentials for a given baryon density. 
The relevant weak interaction processes are the strangeness-changing and flavor-changing decays
\begin{equation}
\begin{split}
    d &\rightarrow u + e^- + \bar{\nu}_e, \\
    s &\rightarrow u + e^- + \bar{\nu}_e.
\end{split}
\end{equation}
Since neutrinos are weakly interacting and are assumed to leave the system without further scattering, their chemical potentials are set to zero.\footnote{Note that during the protoneutron-star phase at early times, this approximation is strongly violated.} 
Under this approximation, the chemical potential conservation relations reduce to
\begin{equation}
    \mu_d = \mu_u + \mu_e, \qquad \mu_s = \mu_d.
\end{equation}
In a bulk system at finite temperature, particle-antiparticle pairs can be thermally created, and the net electric charge must vanish. 
The charge-neutrality condition can be written as
\begin{equation}
    q = \sum_{i = u,d,s,e,\mu^-} n_i q_i = 0,
\end{equation}
where $q_i$ is the electric charge of the particle species $i$. 

Finally, the baryon chemical potential and baryon density are given by
\begin{equation}
    \mu_B := \sum_{f=u,d,s} \mu_f, \quad\rho_B = \frac{1}{3}\sum_{f = u,d,s} n_f,
\end{equation}
The coupled equations of charge neutrality and baryon density are solved simultaneously at finite temperature to determine the chemical potentials for quarks and leptons. These chemical potentials are subsequently employed in the evaluation of the grand canonical potential.

\subsection{Massive Strange Quark Matter in Perturbative QCD}
\begin{figure*}
    \centering
    \includegraphics[width=0.45\linewidth]{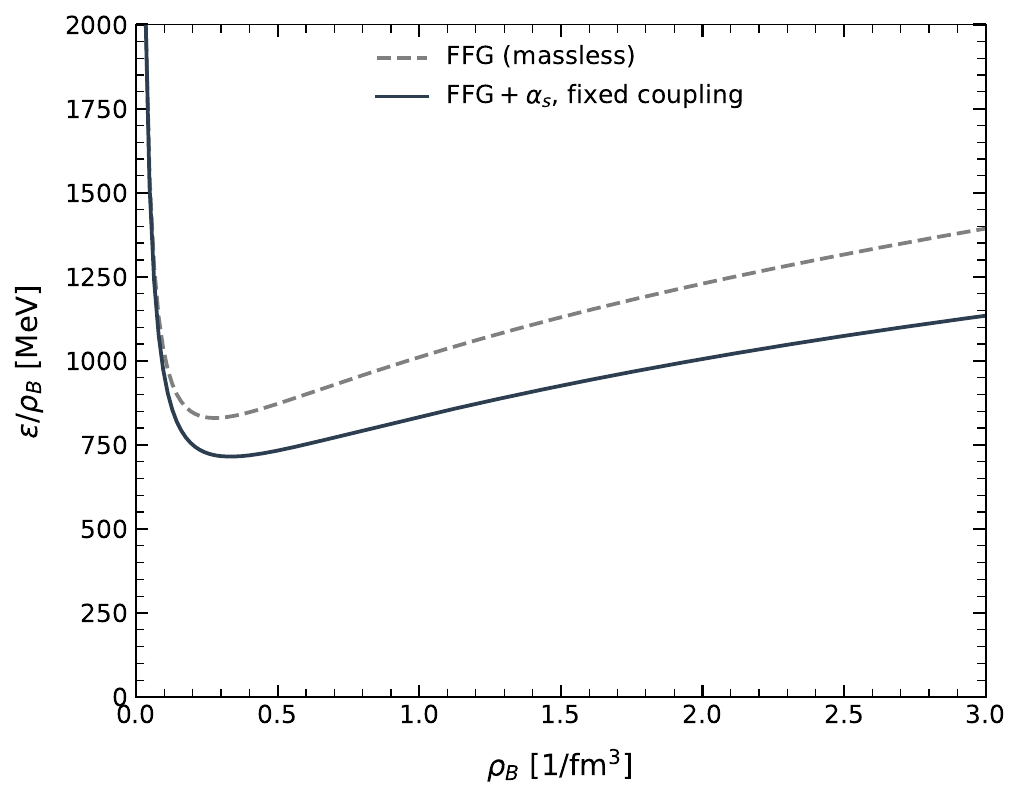}
    \includegraphics[width=0.45\linewidth]{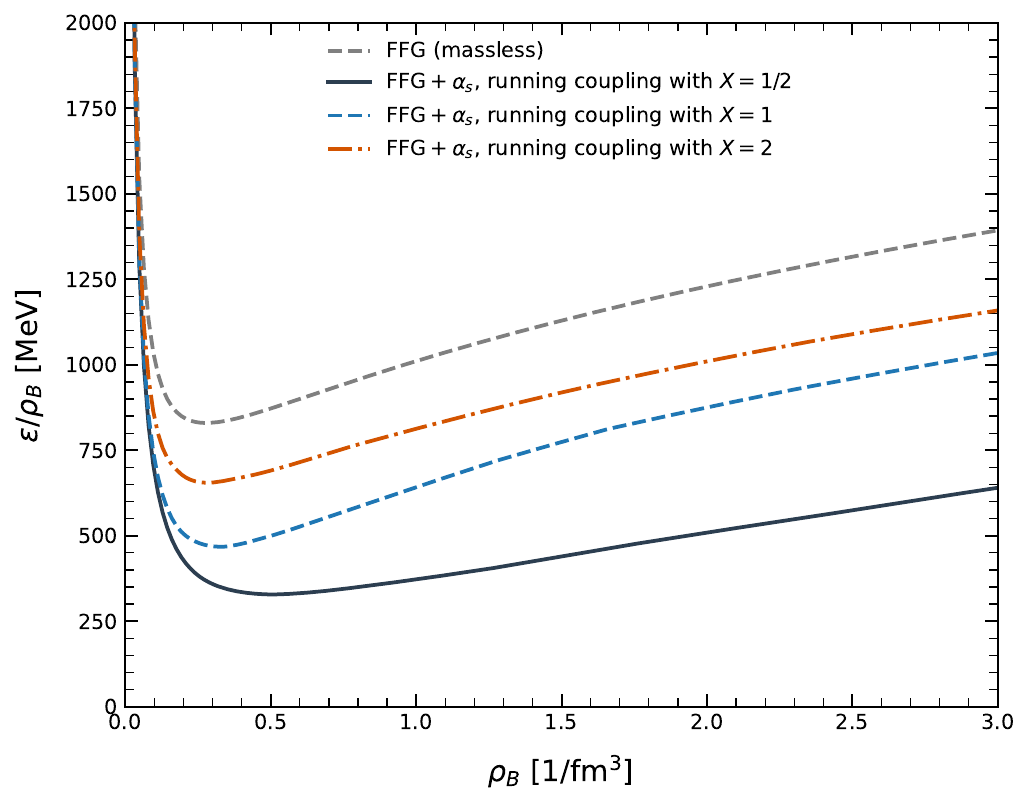}
    \caption{Energy per baryon of strange quark matter as a function of the baryon number density for a fixed QCD coupling (left panel) and a modeled running QCD coupling (right panel). In both panels, we set the bag constant to $B^{1/4}=145~\mathrm{MeV}$ and use the quark masses $m_u=m_d=0$ and $m_s=96~\mathrm{MeV}$.}
    \label{fig:energy_per_baryon}
\end{figure*}

Having established the thermodynamic framework for the massless case, we now present the formulas for the massive case at $\mathcal{O}(\alpha_s)$.
This extension is essential for realistic modeling of SQM, since the
strange-quark mass is comparable to, or larger than, the characteristic
temperature scales encountered in supernovae and compact-binary
mergers.
Following Ref.~\cite{Gorda:2021gha}, the full thermodynamic potential,
including finite quark-mass corrections, can be written as,
\begin{equation}
    \Omega(m_s,\mu,T) = \Omega(m_s=0,\mu,T) + \delta\Omega(m_s,\mu,T),
    \label{eq:grand_potential}
\end{equation}
where the first term on the right-hand side is the grand potential in the massless limit given in Eq.~\eqref{eq:massless potential}, and $\delta\Omega(m_s,\mu,T)$ denotes the approximate correction arising from finite quark masses. 
The analytical expression for $\delta\Omega$ was derived in Ref.~\cite{Gorda:2021gha} and accounts for the leading-order mass corrections to the thermodynamic potential at finite temperature and chemical potential.\footnote{Note that the full numerical result for massive quarks at $\mu$ and $T$ dependence was derived previously in Ref.~\cite{Laine:2006cp}.}
By consistently including both the perturbative QCD corrections and finite quark masses, this framework provides a more realistic and thermodynamically consistent description of SQM, bridging the gap between idealized bag model calculations and full QCD predictions.
The expression for $\delta \Omega$ is as follows
%\begin{widetext}
\begin{align}
\delta\Omega ={}& 
\frac {{N_c} m_s^2 T^2} {12}\left (1 + 12\bar {\mu} _s^2 \right)\nonumber \\
&+\frac {N_c m_s^4} {32\pi^2}\left \{4\ln \!\left(\frac {m_s} {4\pi T} \right) -  2\aleph(z_s ) - 3 \right\} \nonumber\\
&+
\frac {d_A \alpha_s m_s^2 T^2} {48\pi}\biggl \{- 12\bar {\mu} _s\bigl [\bar{\mu}_s +\bar {\mu}_s  \aleph(z_s ) +  4 i\aleph(0,  z_s ) \bigr] \nonumber\\
&\qquad\qquad-  48\aleph(1, z_s ) + 48\ln (A) -  3\aleph(z_s ) - 7 \nonumber\\
&\qquad\qquad+ \left (6 + 72\bar {\mu}_s^2 \right)\ln\!\left (\frac {\Lambda_\text{reno}} {4\pi T} \right) + 6\frac {m_\text{E}} {\pi T} \biggr\}
\end{align}
with $z_s := 1/2 - i \bar{\mu}_s$, $A \approx 1.2824271291006226369$ Glaisher's constant, $\Lambda_\mathrm{reno}$ the renormalization scale in the $\overline{\mathrm{MS}}$ scheme, and with the $\aleph$ functions defined as in Ref.~\cite{Vuorinen:2003fs}:
\begin{align}
    \zeta'(x,y) &:= \partial_x \zeta(x,y), \\
    \aleph(n,z) &:= \zeta'(-n,z)+(-1)^{n+1}\zeta'(-n,z^{*}), \\
    \aleph(z) &:= \frac{\Gamma '(z)}{\Gamma(z)}+\frac{\Gamma '(z^*)}{\Gamma(z^*)} 
\end{align} 
with $\zeta$ the Hurwitz zeta function.

When discussing perturbative corrections to the thermodynamic potential, we consider the running coupling $\alpha_s$ and the strange quark mass dependence. 
As strange stars will probe down to small chemical potentials where the perturbative running of $\alpha_s$ breaks down, in our figures we consider a model for the running coupling where it saturates at small renormalization scales.

Our complete expression for $\alpha_s$ across all three momentum regimes is taken from  Ref.~\cite{Chiba:2023ftg}
\begin{eqnarray}
\alpha_s(Q^2) &=& \alpha_s^{\text{low}}(Q^2)\,\theta(t_{\text{low}}-Q^2) \nonumber\\
&& + \alpha_s^{\text{mid}}(Q^2)\,\theta(Q^2 - t_{\text{low}})\,\theta(t_{\text{high}}-Q^2) \nonumber\\
&& + \alpha_s^{\text{high}}(Q^2)\,\theta(Q^2 - t_{\text{high}})
\label{eq:running_coupling}
\end{eqnarray}
with $\theta$ the Heaviside function, ${t_{\text{low}}^{1/2} := 0.3\;\text{GeV}}$, ${t_{\text{high}}^{1/2} := 1.1\;\text{GeV}}$, ${Q^2 := \Lambda_{\text{reno}}^2}$, and
\begin{equation}
    \Lambda_{\text{reno}} = X \sqrt{(2\,\mu_B/3)^2 + (0.723 \cdot 4\pi T)^2}
\label{eq:temp_eq}
\end{equation}
where $X \in [1/2, 2]$ and $T$ is the temperature in the system~\cite{Kurkela:2009gj,Kurkela:2016was,Gorda:2021gha}.
We also follow Refs.~\cite{Deur:2014qfa,Deur:2016cxb} to take $\alpha_s^\text{low}(Q^2) :=  \alpha_s(0) \exp(Q^2 / 4 \kappa^2)$  with $\alpha_s^\text{low}(0) = 1.22$, $\kappa = 0.51 \,\mathrm{GeV}$.
For the high-momentum domain, we adopt the standard four-loop perturbative QCD running coupling, denoted as $\alpha_s^\text{high}$. 
We set the scale using $\alpha^\text{high}_s\bigr( (2\,\text{GeV})^2\bigr) = 0.2994$~\cite{ParticleDataGroup:2008zun}.
\begin{table*}
\caption{\label{tab:table1}Equilibrium properties of hot-quark stars, namely energy density and energy per baryon at \(T = 5\)~MeV with \(m_u = m_d = 0\) and \(m_s = 96\)~MeV, tabulated as a function of baryon number density. Results are shown for both the free Fermi gas (FFG) and the interacting Fermi gas (FFG + \(\mathcal{O}(\alpha_s)\) perturbative corrections) with a running QCD coupling $(X=1)$ and for various values of the bag constant.}

\begin{ruledtabular}
\begin{tabular}{ccccccc}
 &\multicolumn{3}{c}{  $\text{FFG}+\alpha_s$ fixed coupling }&\multicolumn{3}{c}{  $\text{FFG}+\alpha_s$ running coupling}\\
 $B^{1/4}~$[MeV]&$\rho_B~$[fm$^{-3}$]& $\epsilon~$[MeV/fm$^{3}$]&$\epsilon/\rho_B~$[MeV]&$\rho_B~$[fm$^{-3}$]&$\epsilon~$[MeV/fm$^3$]
&$\epsilon/\rho_B~$[MeV]\\ \hline
 
 135&$0.266$&$178$&$668$&$0.394$ &$172$& $437$\\
 145&$ 0.328$&$235$&$716$&$ 0.468$&$229$ &$ 490$\\
 155&$0.400$&$ 306$&$ 764$&$0.551$&$299$&$543$\\
 
\end{tabular}
\end{ruledtabular}
\end{table*}

The transition between the non-perturbative low-momentum regime and the perturbative high-momentum domain requires a smooth, well-behaved interpolation to avoid unphysical discontinuities in the coupling and its derivatives. 
Such discontinuities would otherwise introduce spurious artifacts in thermodynamic observables, particularly in the speed of sound squared, $c_s^2$, which is highly sensitive to the scale dependence of $\alpha_s$. 
For the intermediate momentum region, we follow Ref.~\cite{Chiba:2023ftg} utilize the running coupling via a degree-five polynomial interpolation, expressed as
\begin{equation}
\alpha_s^{\text{mid}}(Q^2) = \sum_{m=0}^{5} c_m \, \mu_I^m,
\end{equation}
where the six unknown coefficients $c_m$ are uniquely determined by imposing a set of matching conditions at the boundaries between the low, mid, and high regions. 
Specifically, we require that the interpolating function and its first two derivatives are continuous at both transition scales, $Q^2 = t_{\text{low}}$ and $Q^2 = t_{\text{high}}$. 
This ensures a smooth $c_s^2$ connection across the entire momentum range. 
This provides a total of six independent constraints, uniquely fixing all coefficients $c_m$ and guaranteeing a smooth transition without introducing spurious oscillations or kinks that could compromise the reliability of our subsequent EOS calculations.

While the running coupling has been studied for all three momentum regimes (high, mid, and low) as formulated in Eq.~\eqref{eq:running_coupling}, we note that the effect of running quark masses has not been accounted for in our analysis. 
In fact, earlier studies of the EOS of SQM and strange stars at zero temperature also adopted a similar assumption (see Refs.~\cite{Jaffe,Haensel:1986qb,Alcock}).

The temperature adopted in Eq.~\eqref{eq:temp_eq} is motivated by the
finite-temperature conditions relevant to newly formed compact stars.
During the early evolution of a proto-neutron star, neutrinos are
initially trapped because of their short mean free paths and subsequently
escape through diffusion as the star deleptonizes and cools. The average
energies of neutrinos and antineutrinos emitted during the cooling phase
of a core-collapse supernova are typically of order
$10$--$20~\mathrm{MeV}$~\cite{SN1987A,Beacom:2010kk}. For a
Fermi--Dirac spectrum with vanishing chemical potential, the average
energy is related to the effective neutrino temperature by
$\langle E_\nu\rangle \simeq 3.15\,T_\nu$. Motivated by this characteristic energy scale, we adopt $T=5~\mathrm{MeV}$ as a reference matter temperature for studying
finite-temperature SQM and possible early-stage strange-star
configurations. We present the corresponding EOS and stability analysis
for massive SQM. Results at other temperatures can be obtained within
the same framework.

In Fig.~\ref{fig:pres1}, we show the pressure of SQM, normalized to the
free-Fermi-gas pressure, as a function of the baryon chemical potential
for fixed and running strong couplings. We take the bag constant to be
$B^{1/4}=145~\mathrm{MeV}$. For the fixed-coupling prescription,
$P/P_{\mathrm{FFG}}$ rises rapidly and approaches an approximately
constant value of $0.8$ at large $\mu_B$. The running-coupling results
exhibit a stronger dependence on both $\mu_B$ and the
renormalization-scale parameter $X$. In particular, increasing $X$
raises the pressure and shifts the zero-pressure point toward a lower
baryon chemical potential. The zero-pressure condition, which
determines the surface of self-bound SQM, occurs at approximately
$\mu_B\simeq868~\mathrm{MeV}$ in the fixed-coupling case, compared with
$\mu_B\simeq1112$, $983$, and $883~\mathrm{MeV}$ in the
running-coupling cases with $X=1/2$, $1$, and $2$, respectively. Thus,
relative to the fixed-coupling result, the running-coupling prescription
shifts the zero-pressure point toward larger $\mu_B$ for all three
values of $X$ considered, although the magnitude of this shift depends
strongly on $X$.

In Fig.~\ref{fig:energy_per_baryon}, we show the energy per baryon as a
function of the baryon density. We present results for SQM in three
cases: a free Fermi gas with confinement implemented through the bag
constant, the $\mathcal{O}(\alpha_s)$ result with a fixed coupling,
$\alpha_s\approx0.3$, and the $\mathcal{O}(\alpha_s)$ result with a
running coupling. In the fixed-coupling case, the minimum energy per
baryon occurs at a lower baryon density, whereas in the
running-coupling case, the minimum is shifted toward a higher baryon density.
The perturbative corrections appreciably reduce the minimum energy per
baryon, indicating an increased stability of quark matter. Furthermore,
for both coupling prescriptions, the energy per baryon at zero pressure
remains below $930.4~\mathrm{MeV}$, satisfying the conventional
criterion for the absolute stability of SQM relative to ordinary
nuclear matter.

For practical applications, we present in Table~\ref{tab:table1} the equilibrium properties of three representative EOSs, corresponding to the bag constant values $B^{1/4} = 135,\,145,\,155\,\text{MeV}$. 
These EOSs all remain causal, and each supports compact stars of more than $1.4\,M_\odot$. 
We note that our chosen values of bag constant lie within the ballpark of stability suggested by perturbative QCD with the bag parameter fitted to lattice-QCD simulations of isospin-dense matter at zero temperature; the same analyses also indicate an upper bound $B^{1/4} \lesssim 160\,\text{MeV}$~\cite{Bai:2024amm}.

 %TG here on final read-thru

\begin{table*}
\caption{\label{tab:table2}
Radius, compactness, and tidal deformability of canonical $1.4\,M_\odot$ strange quark stars at $T=5$~MeV, with $m_u=m_d=0$ and $m_s=96$~MeV, for different values of the bag constant $B$, including perturbative QCD corrections with fixed and running coupling with $X=1$.}
\begin{ruledtabular}
\begin{tabular}{ccccccc}
 &\multicolumn{3}{c}{  $\text{FFG}+\alpha_s$ fixed coupling }&\multicolumn{3}{c}{  $\text{FFG}+\alpha_s$ running coupling}\\
 $B^{1/4}~$[MeV]&$R~$[km]& $C~$&$\Lambda~$&$R~$[km]&$C~$&$\Lambda$ \\ \hline
 
 135&$12.07$
 &$0.171$& $110.4$&$12.13$&$0.170$&$104.5$\\
  145&$10.85$
 &$0.190$& $64.87$&$10.90$&$0.189$&$67.11$\\
  155&$9.77$
 &$0.211$& $35.95$&$9.81$&$0.210$&$37.89$\\
 
\end{tabular}
\end{ruledtabular}
\end{table*}

\begin{figure*}[htp]
\setkeys{Gin}{width=0.45\linewidth}
    %\subfloat[caption 1]
    {\includegraphics{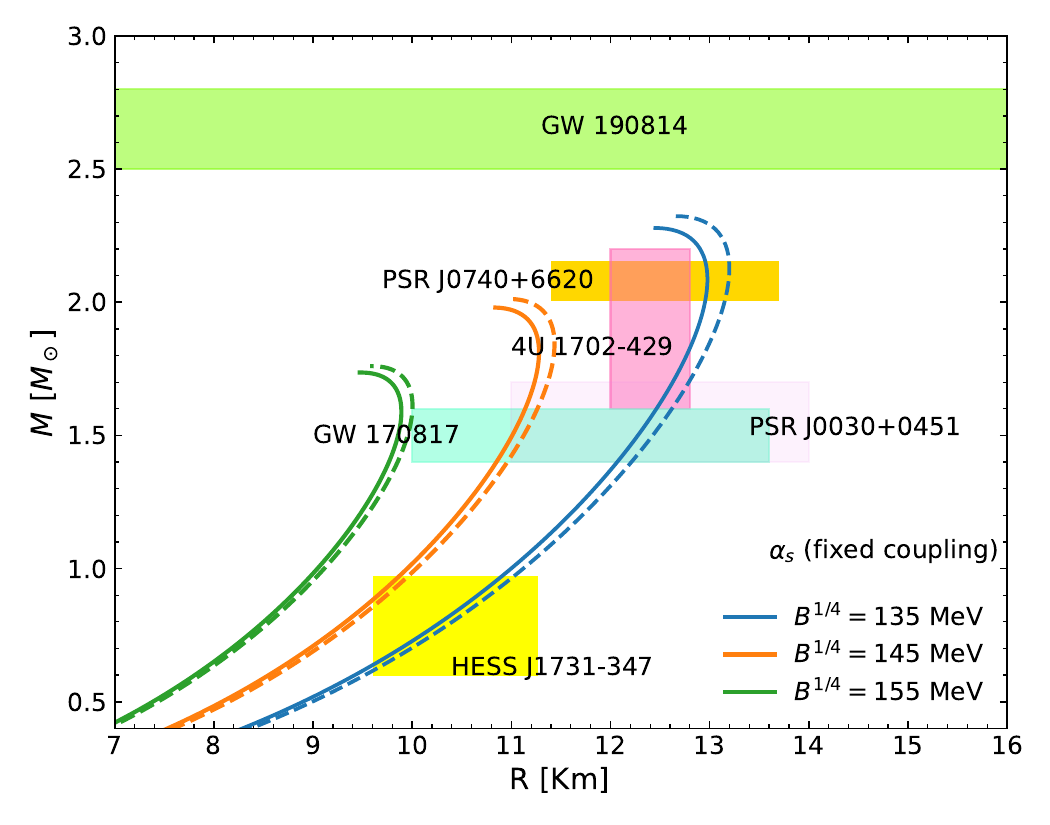}}
    %\subfloat[caption 2]
    {\includegraphics{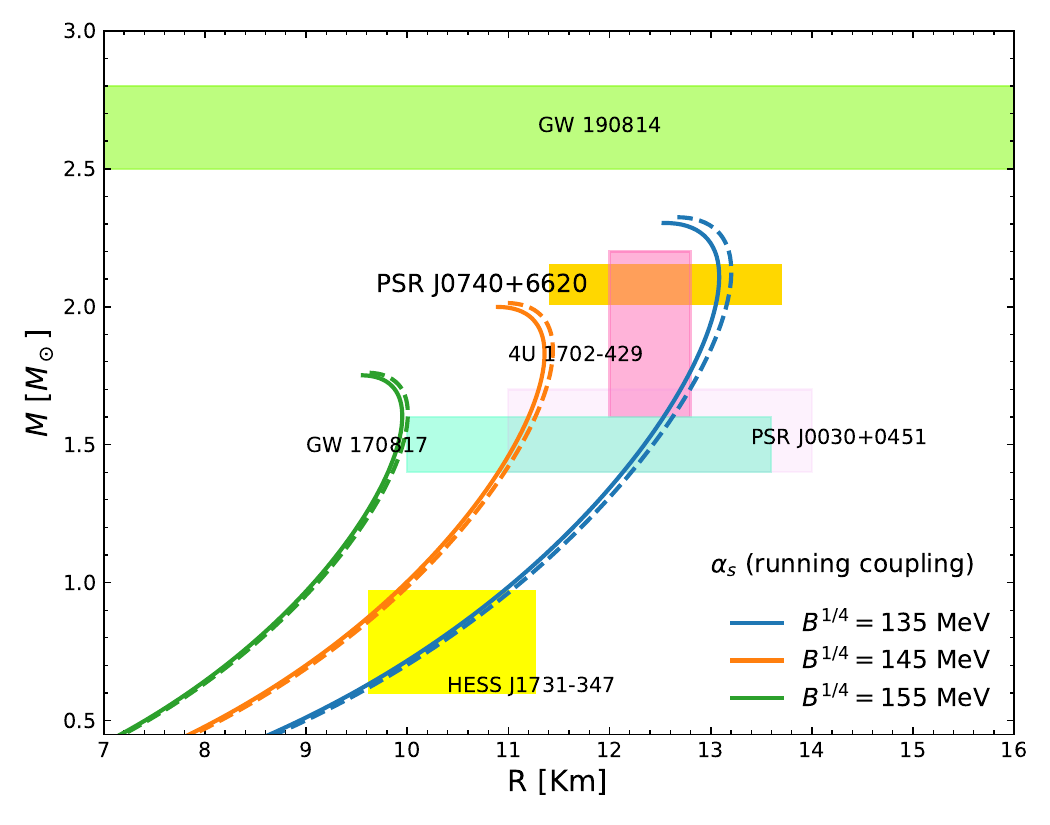}}

\caption{Mass--radius relations for strange stars obtained using a fixed coupling, $\alpha_s \simeq 0.3$ (left panel), and a running coupling with $X=1$ (right panel), at a temperature of $T=5~\mathrm{MeV}$. 
Solid curves correspond to $m_u=m_d=0$ and $m_s=96~\mathrm{MeV}$, whereas dashed curves represent the massless-quark case, $m_u=m_d=m_s=0$. 
Different colors indicate bag constants in the range $B^{1/4}=135$--$155~\mathrm{MeV}$. 
The cyan and green bands show observational constraints on the radius of a $1.4\,M_\odot$ compact star from GW170817~\cite{LIGOScientific:2017vwq} and on the mass of the secondary compact object in GW190814~\cite{LIGOScientific:2020zkf}. 
The light-pink region represents the mass--radius constraints for
PSR J0030+0451~\cite{Vinciguerra:2023qxq}, whereas the dark-yellow region represents those of PSR J0740+6620~\cite{Miller:2021qha}, based on observations from NICER and XMM-Newton. 
The horizontal black line represents the mass measurement of PSR J0348+0432~\cite{Antoniadis:2013pzd}. 
The yellow region indicates constraints associated with the central compact object in the supernova remnant HESS J1731--347~\cite{2022NatAs...6.1444D,Horvath:2023uwl}, and the dark-pink region represents the constraint on the low-mass X-ray binary 4U 1702$-$429 inferred from Rossi X-ray Timing Explorer observations~\cite{Nattila:2017wtj}.}

\label{fig:MR_curve}
\end{figure*}

\section{Global Properties of Quark Stars}
\label{sec:Global Properties of Quark Stars}

We study the hydrostatic equilibrium properties of nonrotating strange stars, by solving the Tolman-Oppenheimer-Volkoff (TOV) equations~\cite{PhysRev.55.364,PhysRev.55.374}. 
To solve TOV equations, we adopt the relativistic enthalpy formulation of Lindblom~\cite{Lindblom:2013kra}, which allows us to determine the mass \(m(h)\) and radius \(r(h)\) as functions of the enthalpy \(h\) in a numerically efficient manner. 
Additionally, we compute the tidal deformability \(\Lambda\) and compactness \(C\) for strange stars thereby establishing the full set of macroscopic observables needed to confront our EOS predictions with astrophysical observations~\cite{Hinderer:2009ca}.

In Fig.~\ref{fig:MR_curve}, we present the mass--radius relations for strange stars computed from our perturbatively corrected EOS for three representative bag constants, \(B^{1/4} = 135\), \(145\), and \(155\) MeV, while consistently incorporating finite strange-quark masses. 
We observe that, for higher values of the bag constant, $B^{1/4}= 155$~MeV, the resulting configurations are more compact and their masses lie below $2\,M_\odot$, whereas lower values of the bag constant allow for more massive configurations. For both the fixed- and running-coupling prescriptions, our EOS supports
strange stars with maximum masses of approximately $2.35\,M_\odot$. 

We see from Fig.~\ref{fig:MR_curve} that our EOSs are in good agreement with existing multi-messenger constraints~\cite{Antoniadis:2013pzd,Miller:2021qha,Vinciguerra:2023qxq,2022NatAs...6.1444D,Horvath:2023uwl} (the shaded regions). 
For a canonical $1.4\,M_\odot$ strange star, the predicted radii fall in the range of $9$ to $13$~km (see Table~\ref{tab:table2}), while a $2\,M_\odot$ strange star corresponds to radii of approximately $11$ to $13$~km.

\begin{figure}[t]
    \centering
    % Left image
    \includegraphics[width=0.9\linewidth]{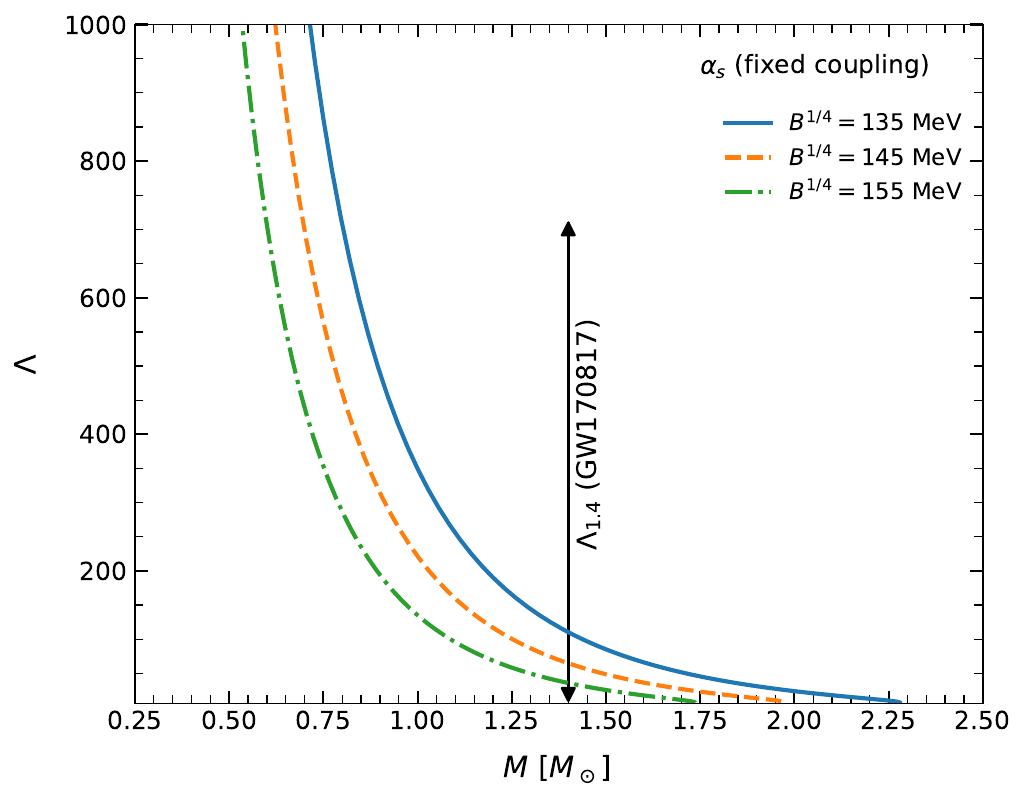}
    \hfill
    % Right image
    \includegraphics[width=0.9\linewidth]{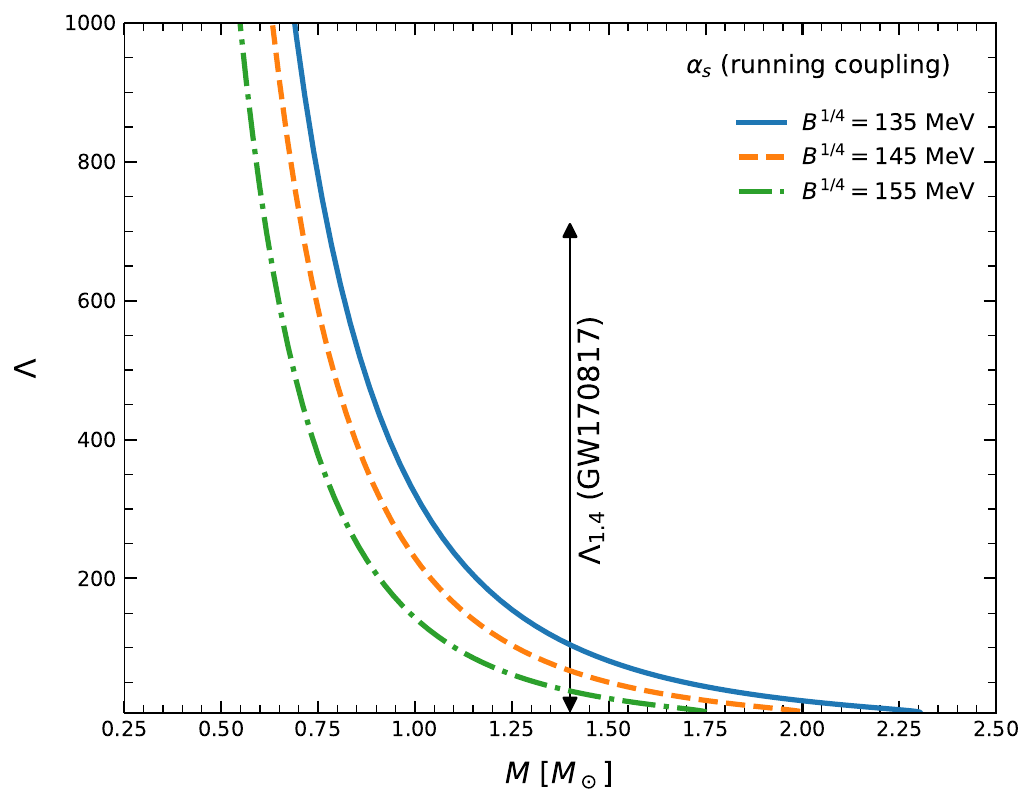}

    \caption{Mass--tidal-deformability relations for strange stars at
$T=5~\mathrm{MeV}$, with $m_u=m_d=0$ and $m_s=96~\mathrm{MeV}$. The
colored curves correspond to bag constants in the range
$B^{1/4}=135$--$155~\mathrm{MeV}$. The vertical error bar indicates the
tidal-deformability constraint from GW170817~\cite{LIGOScientific:2017vwq},
obtained assuming a low-spin prior. The upper and lower panels show the
results for a fixed coupling, $\alpha_s\simeq0.3$, and a running QCD
coupling with $X=1$, respectively.}

    \label{fig:combined_tidal}
\end{figure}

In Fig.~\ref{fig:combined_tidal}, we present the mass--tidal-deformability relation for strange quark stars derived from our perturbatively corrected EOS, incorporating a running QCD coupling and finite strange quark masses. 
We see from this figure that all three EOS curves fall well within the \(90\%\) credible region inferred from the GW170817 binary-neutron-star merger under the low-spin prior~\cite{LIGOScientific:2017vwq}. 
Furthermore, we observe a clear inverse correlation between the bag constant and the tidal deformability: smaller \(B^{1/4}\) yields larger radii and less compact stars, which are more readily deformed by the companion's tidal field, resulting in larger \(\Lambda\). 

\section{Conclusion}
\label{sec:conclusion}
We have presented a comprehensive thermodynamic treatment of
finite-temperature strange quark matter (SQM), incorporating
perturbative QCD (pQCD) corrections through
$\mathcal{O}(\alpha_s)$ for both fixed and running strong couplings.
These corrections modify the energy density and pressure of massive
SQM, thereby significantly affecting the equilibrium configurations of
proto-quark stars that may form in core-collapse supernovae. For the
model parameters considered, the running-coupling prescription
generally produces more stable configurations. Moreover, both coupling
prescriptions yield zero-pressure energies per baryon below that of
the most stable iron nucleus, satisfying the absolute-stability
criterion for SQM and supporting the possibility that SQM constitutes
the ground state of strongly interacting matter.

Utilizing the presented thermodynamically consistent equation of state (EOS), we investigated the bulk properties of strange-star candidates, ensuring compatibility with empirical constraints on mass, radius, and tidal deformability derived from pulsar and LIGO gravitational-wave observations. The resulting mass--radius and mass--tidal-deformability correlations serve as a crucial bridge connecting the underlying microphysics of dense, strongly interacting matter to macroscopic astrophysical data, encompassing radio-pulsar mass measurements, radius inferences from NICER and XMM-Newton, and the gravitational-wave signatures of binary mergers such as GW170817. 

% TG HERE
Moving forward, next-generation telescopes will provide a wealth of new observational data about the dense-matter EOS, and theorists will need to ensure that EOS inference incorporates as much theoretical information as possible, including pQCD input \cite{Gorda:2022jvk,Komoltsev:2023zor,Gorda:2025aiu,Gorda:2026rzm}. 
Such pQCD-informed EOS inference not only sharpens high-density EOS bounds but also improves the prospects for identifying potential strange quark stars in future multi-messenger campaigns. 
%(see Ref.~\cite{Gangopadhyay:2013gha}). 
Additionally, post-merger gravitational-wave signals from binary neutron star coalescences would provide direct access to the finite-temperature regime of the EOS, potentially revealing information about the composition and interaction channels of dense matter that is not encoded in the zero-temperature EOS. 
For example, transport signatures of deconfined quark phases may offer a promising way to probe the presence of quark matter in the hot remnants of such mergers~\cite{CruzRojas:2024etx,Hernandez:2024hcl,Hernandez:2025zxw}. 
The upcoming generation of gravitational-wave observatories is expected to reach the sensitivity needed to directly probe these finite-temperature EOS effects during both the merger and post-merger phases~\cite{Raithel:2023gct,ET:2025xjr}.

\begin{acknowledgments}

TR acknowledges Aleksi Vuorinen, Oleg Korobkin, Soumi Dey, Rahul Somasundaram, and the Los Alamos National Laboratory for support and hospitality.

\end{acknowledgments}

%\nocite{*}

\bibliography{apssamp}% Produces the bibliography via BibTeX.

\end{document}